\documentstyle[12pt,epsf]{article}
\newcommand{\tdm}[1]{\mbox{\boldmath $#1$}}

\begin{document}
\begin{titlepage}
\pagestyle{empty}
\vspace*{2cm}
\begin{center}
{\large\bf  Possible probe of the QCD odderon singularity\\
through the quasidiffractive} $\eta_c$ {\large\bf production in}
$\gamma \gamma$ {\large\bf collisions .}
\vspace{1.1cm}\\
         {\sc L.~Motyka}$^a$,
         {\sc J.~Kwieci\'nski}$^b$,
\vspace{0.3cm}\\
$^a${\it Institute of Physics, Jagellonian University,
Cracow, Poland}
\vspace{0.3cm}\\
$^b${\it Department of Theoretical Physics, \\
H.~Niewodnicza\'nski Institute of Nuclear Physics,
Cracow, Poland}
\end{center}
\vspace{1.5cm}
\begin{abstract}
The reactions $\gamma \gamma \rightarrow \eta_c \eta_c$ and
$\gamma \gamma \rightarrow \eta_c + X$ are discussed within the
three gluon exchange model.  We give predictions for the differential
cross-sections and discuss feasibility of measuring these processes at LEP2
and TESLA.  The total cross-sections were estimated to be approximately
equal to 40 fb
and 120 fb for $\gamma \gamma \rightarrow \eta_c \eta_c$ and
$\gamma \gamma \rightarrow \eta_c + X$ respectively assuming exchange
of elementary gluons that corresponds to the odderon with intercept
equal to unity. These values can be enhanced by a factor equal to
1.9 and 2.1 for LEP2 and TESLA energies if the odderon intercept
is equal to 1.07.
The estimate of cross-sections
$\sigma (e^+ e^- \rightarrow e^+ e^- \eta_c \eta_c) $ and
$\sigma (e^+ e^- \rightarrow e^+ e^- \eta_c  X) $ for untagged $e^+$ and
$e^-$ is also given.
\end{abstract}
\vspace{1cm}

\noindent
{\sf TPJU-1/98}\\
{\sf February 1998}
\end{titlepage}

The dominant contribution in the high-energy limit of perturbative QCD
is given by the exchange of interacting gluons \cite{LIPATOV1,LIPATOV2}.
The fact that the gluons have spin equal to unity automatically implies
that the cross-sections corresponding to the exchange of elementary gluons
are independent of the incident energies while the interaction between gluons
leads to increasing cross-section.    Besides the pomeron
\cite{LIPATOV1,LIPATOV2,LIPATOV3,BFKL,GLR}
one also expects presence of the so called ``odderon" singularity
\cite{LIPATOV1,LIPATOV2,LIPATOV3,BARTELS,KP}.
In the leading logarithmic
approximation the pomeron is described by the BFKL equation which corresponds
to the sum of ladder diagrams with the reggeized gluon exchange along the 
ladder. The odderon is described by the three-gluon
exchange.  Unlike pomeron which corresponds to the vacuum quantum numbers
and so to  the positive charge conjugation
the  odderon is characterised by $C=-1$ (and I=0)  i.e.\ it carries
the same quantum
numbers as the $\omega$ Regge pole.  The (phenomenologically) determined
intercept $\lambda_{\omega}$ of the $\omega$ Regge pole is
approximately equal to $1/2$
\cite{DOLA}.  The novel feature of the odderon singularity corresponding
to the gluonic degrees of freedom is the potentially very high value of its
intercept $\lambda_{odd}\gg\lambda_{\omega}$.  The exchange of the
three (noninteracting) gluons alone generates singularity with intercept
equal to unity while interaction between gluons described in the leading
logarithmic
approximation by the BKP equation \cite{BARTELS,KP,JW} is even capable to boost
the odderon intercept above unity \cite{GLN} \footnote{The results of ref.
\cite{GLN} are however in conflict with recent analysis \cite{BRAUN}}. 
The energy dependence of the amplitudes corresponding to $C=-1$ exchange
becomes similar to the diffractive ones which are controlled by the
pomeron exchange.

Possible tests of both QCD perturbative pomeron as well as of odderon have
to rely on (semi) hard processes where presence of hard scale can justify
the use of perturbative QCD.
   Very useful measurement in this respect
 is the very high
 energy   exclusive
 photo (or electro) production of heavy charmonia (i.e. $J/\Psi$
 \cite{RYSKIN,BRODSKY} or
 $\eta_c$ \cite{MSN,CKMS,IKS} etc.  for probing the QCD pomeron or odderon
 respectively).  The estimate of the odderon contribution to the
 photo- (or electro-) production of even charge conjugation mesons does however
 require model assumptions about the coupling of the three gluon
 system to a proton.  It would therefore be useful to consider the
 process which could in principle be calculated entirely within
 perturbative QCD.  The relevant measurement which fulfills those criteria is the exclusive
 quasidiffractive
 production of even $C$ charmonia in $\gamma \gamma$ collisions or, to be
 precise, the processes: $\gamma \gamma \rightarrow \eta_c \eta_c$ or
  $\gamma \gamma \rightarrow \eta_c + X $\cite{GINZBURG,ODDRUS}.
The main purpose of our paper is to present the  theoretical and phenomenological
description of the double $\eta_c$ production in high energy $\gamma \gamma$
collisions and of the process $\gamma \gamma \rightarrow \eta_c + X$ with
the large rapidity gap
between $\eta_c$ and the hadronic system $X$ assuming the
three gluon exchange mechanism.  In our paper we shall follow the formalism
developed in ref \cite{ODDRUS} where the production of
pseudosclar mesons in $\gamma \gamma$ collisions within the three gluon
exchange mechanism is discussed with great detail.

\noindent
\begin{figure}
\hbox{
\epsfxsize = 15.5cm
\epsfysize = 7.3cm 
\epsfbox{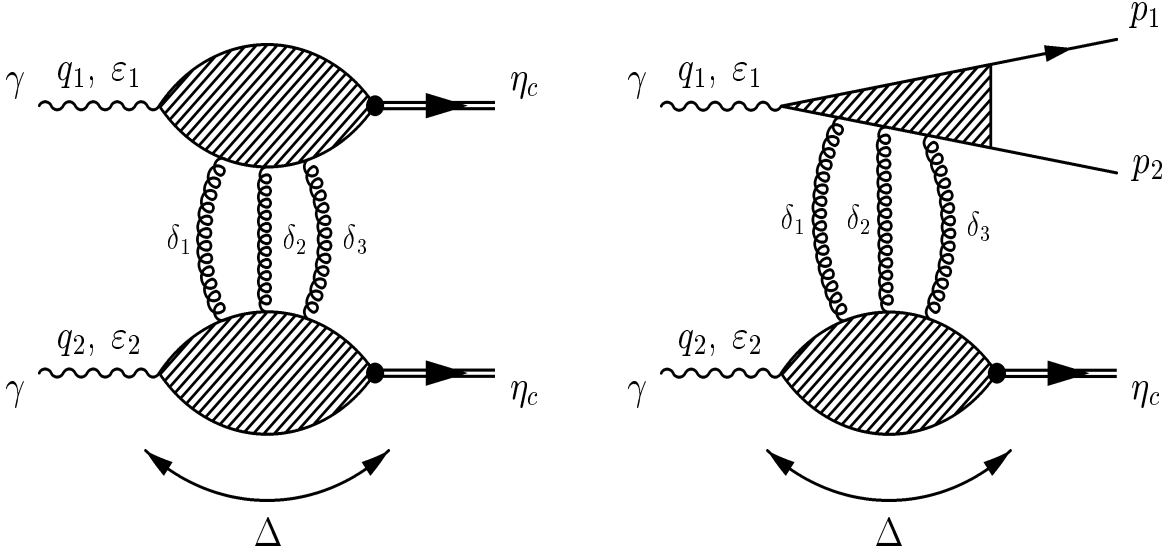} \vspace{0.5cm}
}
\caption{Kinematics of the three gluon exchange mechanism of the process 
$\gamma\gamma \rightarrow \eta_c \eta_c $ (1a) and 
$\gamma\gamma \rightarrow \eta_c + X$ (1b).} 
\end{figure}

The kinematics of the three gluon exchange diagram to the
processes:
$\gamma^{*}(q_1) + \gamma^{*}(q_2) \rightarrow \eta_c \eta_c$ and
$\gamma^{*}(q_1) + \gamma^{*}(q_2) \rightarrow \eta_c + X$
 is illustrated in
Fig.~1a and Fig. ~1b.
%

The amplitude $M^{ij}$ for
the process $\gamma^{*}(q_1) + \gamma^{*}(q_2) \rightarrow \eta_c \eta_c$
which corresponds to the transverse polarisation of both photons
can be written in the following way:
\begin{equation}
M^{ij}=W^2 C_{qq} \int { d^2\tdm\delta_{1} d^2\tdm\delta_{2} \over
\tdm\delta_{1}^2\tdm\delta_{2}^2\tdm\delta_{3}^2 }
\Phi^i_{\gamma \eta_c}(Q_1^2,\tdm\delta_{1}, \tdm\delta_{2}, \tdm\Delta )
\Phi^j_{\gamma \eta_c}(Q_2^2,-\tdm\delta_{1}, - \tdm\delta_{2}, -\tdm\Delta )
\label{mij}
\end{equation}
where $i$ and $j$ are the polarisation indices and $C_{qq}$ is given by:
\begin{equation}
C_{qq}={10\over 9 \pi N_c^2} \alpha_s^3(m_c^2)
\label{cqq}
\end{equation}
and $\tdm\Delta $ and $\tdm\delta_{i}$ denote the transverse components
of $\Delta$ and $\delta_i$ while $Q_i^2 = -q_i^2$.

The relevant diagrams describing
the impact factor $\Phi^{i,j}_{\gamma \eta_c}$ are given in Fig.~2.
Besides the diagrams presented in Fig.~2 one has also to include
those with the reversed direction of the quark lines.
In the nonrelativistic approximation one gets the following expression
for $\Phi^{i}_{\gamma \eta_c}$:
\begin{equation}
\Phi^{i}_{\gamma \eta_c}=
F_{\eta_c} \sum_{k=1}^2 \epsilon_{ik} \left[
{\tdm\Delta  ^k\over \bar M_1^2+ \tdm\Delta ^2} +
\sum_{s=1}^3 {2\tdm\delta_{s} ^k - \tdm\Delta^k  \over
             \bar M_1^2+(2\tdm\delta_{s} - \tdm\Delta )^2}
                                     \right]
\label{impfi}
\end{equation}
where
\begin{equation}
F_{\eta_c} =\sqrt{
    {4m_{\eta_c} \Gamma_{\eta_c \rightarrow \gamma \gamma} \over
\alpha q_c^2}}
\label{fetac}
\end{equation}
and
\begin{equation}
\bar M_1^2=4m_c^2 + Q_1^2 .
\label{barm12}
\end{equation}
In equations (\ref{impfi},\ref{fetac},\ref{barm12})
$m_c, m_{\eta_c}, \Gamma_{\eta_c \rightarrow \gamma \gamma}$ and $q_c$
denote the charmed quark mass, the mass of the $\eta_c$, the $\eta_c$
radiative width and the charge of the charm quark respectively.  The formula
for the impact factor $\Phi^j_{\gamma \eta_c}$ corresponding to the lower
vertex is given by equation (\ref{impfi}) after changing the polarisation
index $i$ into $j$ and after reversing the sign of $\tdm\delta_{s}$ and of
$\tdm\Delta $ in this equation and after changing $\bar M_1^2$ into
$\bar M_2^2$  given by equation (\ref{barm12}) with $Q_2^2$ instead
of $Q_1^2$. The approximations leading to the formula (\ref{impfi}) are
discussed in \cite{CKMS}.

\noindent
\begin{figure}
\hbox{
\epsfxsize = 15.5cm
\epsfysize = 9.3cm 
\epsfbox{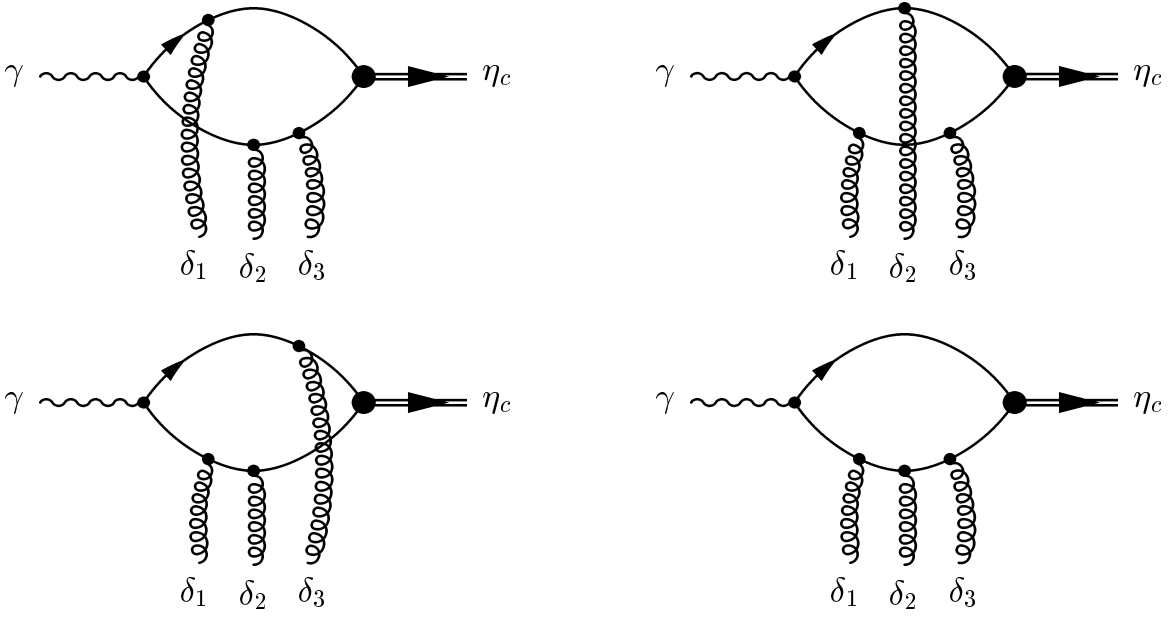}
}
\caption{The diagrams defining the impact factor $\Phi_{\gamma\eta_c}$.}
\end{figure}

For $Q_1 ^2 = Q_2 ^2 \simeq 0$ it is convenient to represent the amplitude
$M^{ij}$ in the following way:
\begin{equation}
M^{ij}=W^2\left(M_1 g^{ij} +
                M_2{\tdm\Delta ^i \tdm\Delta ^j\over \tdm\Delta ^2}\right)
\label{mijdec}
\end{equation}
The corresponding formula for the differential cross-section averaged
over the transverse photons polarizations for the process
$\gamma \gamma \rightarrow \eta_c \eta_c$
reads:
\begin{equation}
{d\sigma \over dt} = {1\over 64 \pi} [(M_1+M_2)^2 + M_1^2]
\label{dsdt}
\end{equation}
For real $\gamma-s$ we have, of course, set $Q_{1,2}^2=0$.

\noindent
\begin{figure}
\hbox{
\epsfxsize = 15.5cm
\epsfysize = 9.3cm 
\epsfbox{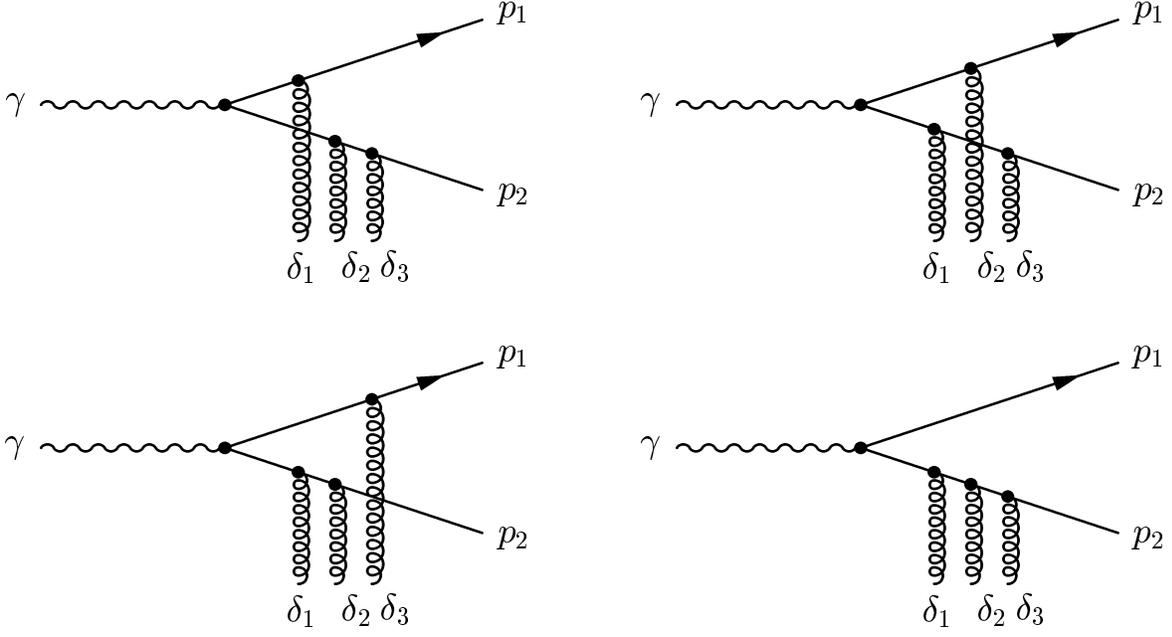}
}
\caption{The diagrams defining the impact factor $\Phi_{\gamma X}$.}
\end{figure}

The process $\gamma \gamma \rightarrow \eta_c + X$ is given by
the diagram of Fig.~1b and the amplitude which corresponds to this diagram
can be written as below:

\begin{equation}
M_{\gamma \gamma \rightarrow \eta_c + X} = W^2 \tilde C_{qq}
\int {d^2\tdm\delta_{1}d^2\tdm\delta_{2}\over
\tdm\delta_{1}^2 \tdm\delta_{2}^2 \tdm\delta_{3}^2}
\Phi_{\gamma \eta_c}(Q_2^2,-\tdm\delta_{1}, -\tdm\delta_{2}, -\tdm\Delta )
\Phi_{\gamma X}(Q_1^2,\tdm\delta_{1}, \tdm\delta_{2}, \tdm\Delta ,p_1,p_2)
\label{metax}
\end{equation}
where
\begin{equation}
\Phi_{\gamma \eta_c}= \sum_{i,j=1}^2\epsilon_{2}^i
\epsilon_{ij}\Phi^j_{\gamma \eta_c}
\label{phirel}
\end{equation}
and
\begin{equation}
\tilde C_{qq} = {10 \over 9\pi N_c ^2}[\alpha_s (m_c ^2) \tilde\alpha_s]^{3/2}.
\label{cqqbar}
\end{equation}
The coupling constant $\tilde\alpha_s$ is defined by the hard scale
characteristic for the upper vertex in Fig.~1b.
The impact factor $\Phi_{\gamma X}$ is given by the following equation:
\begin{equation}
\Phi_{\gamma X}(Q_1^2,\tdm\delta_{1}, \tdm\delta_{2},
\tdm\Delta , p_1, p_2)=
 -ieq_f \bar u(p_1)[m_f R \epsilon_1\hspace{-0.7em}/\,
                  + 2z\bar Q \epsilon_1 +
                  {\bar Q}\hspace{-0.7em}/\,
                  \epsilon_1\hspace{-0.7em}/\,]
                  q_2\hspace{-0.7em}/\; v(p_2)
\label{phietax}
\end{equation}
where
\begin{equation}
\tdm{\bar Q} = \left[{\tdm p_{1}\over m_f^2 + \tdm p_{1}^2} + \sum_{s=1}^3
{\tdm\delta_{s}-\tdm p_{1}\over m_f^2 +
  (\tdm\delta_{s}- \tdm p_{1})^2}\right]
+(\tdm p_{1} \rightarrow \tdm p_{2})
\label{qbar}
\end{equation}
and
\begin{equation}
 R = \left[{-1\over m_f^2 + \tdm p_{1}^2} + \sum_{s=1}^3
{1\over m_f^2 +(\tdm\delta_{s}- \tdm p_{1})^2}\right]
+(\tdm p_1 \rightarrow \tdm p_2).
\label{r}
\end{equation}
In equations (\ref{phietax}, \ref{qbar},\ref{r}) $\bar Q$ is a four-vector
with transverse components $\tdm{\bar Q}$ and vanishing longitudinal
componenets, $m_f$ denotes the mass of the (light) quark
produced as the pair of $q \bar q$ jets, $q_f$ its charge while $z$ is the
component of the photon four momentum $q_1$ carried by a quark jet.
The four momenta $p_1$ and $p_2$ denote the four momenta of the quark
(antiquark) in the final state and $\tdm p_1$, $\tdm p_2$ are their
transverse parts respectively.
When calculating the cross-sections we shall make the
"equivalent quark approximation" \cite{ODDRUS,GINZVM}
which corresponds to the approximation of setting $m_f$ equal to zero
and to retaining the dominant term in $\tdm{\bar Q}$ i.e.:
\begin{equation}
\tdm{\bar Q} = {\tdm p_{1}\over \tdm p_{1}^2} +
               {\tdm p_{2}\over \tdm p_{2}^2}.
\label{eqqa}
\end{equation}

The remaining terms in (\ref{qbar}) can also be large for $\tdm\delta_s \simeq
\tdm p_i$ but their dependence on $\tdm\delta_s$ is such that after the
integration over $\tdm\delta_s$ performed in (\ref{metax}) they are
suppressed in comparison to the leading terms.

The differential cross-section for the
process $\gamma\gamma \rightarrow \eta_c + X(q\bar q)$  
averaged over photon polarisations is given by the
following formula:
\begin{equation}
d\sigma = {(2\pi)^4 \over 2 W^2}\;  \sum_{\{ ... \} }
M^+ _{\gamma\gamma \rightarrow \eta_c + X}
M _{\gamma\gamma \rightarrow \eta_c + X}   \;
dPS_3 (\gamma\gamma \rightarrow \eta_c + X(q\bar q))
\label{dsigetax}
\end{equation}
where $\{ ... \}$ stands for incident photons helicities, outgoing light
quark colours, flavours and polarisations and
$dPS_3 (\gamma\gamma \rightarrow \eta_c + X(q\bar q))$
is the standard parametrisation of the Lorentz invariant tree-body
phase space. The decomposition property
\begin{equation}
dPS_3 (\gamma\gamma \rightarrow \eta_c + X(q\bar q)) =
(2\pi)^3\; dM_X ^2 \; dPS_2(\gamma\gamma \rightarrow \eta_c + X)
                   \; dPS_2 (X \rightarrow q\bar q)
\label{decompps}
\end{equation}
is employed and integration over invariant mass squared $M_X ^2$ of
the $q\bar q$ system is performed.
The remaining integration over the two body phase space
$ dPS_2 (X \rightarrow q\bar q)$ in the applied  approximation can not
be extended to the regions $\tdm p_i ^2 \simeq 0$ since  the integral
would then be divergent. We do therefore introduce, following
ref.~\cite{ODDRUS,GINZVM}
a physical cut-off $\mu$  which is defined by
the light quarks constituent masses and we set for its magnitude 
$\mu = 0.3\,{\rm GeV}$. The integral is bounded  from above by
the condition $\tdm p_i ^2 < \tdm\Delta ^2$ which assures the diffractive
nature of the process. Thus we obtain a logarithmic expression
${\rm ln\;} |t| / \mu ^2$ arising from the integral 
$\int_{\mu ^2} ^{|t|} d\tdm p_1 ^2 / \tdm p_1 ^2$.
Finally the differential cross-section reads
\begin{equation}
{d\sigma \over dt} = {1 \over 64\pi} \;
\left| \tilde C_{qq} \int {d^2\tdm\delta_{1}d^2\tdm\delta_{2}\over
\tdm\delta_{1}^2 \tdm\delta_{2}^2 \tdm\delta_{3}^2}
\Phi_{\gamma \eta_c}(Q_2^2,-\tdm\delta_{1}, -\tdm\delta_{2}, -\tdm\Delta ,p_1,p_2)
\right|^2 \; {2\alpha \bar e ^2 \over 3\pi}
\; {\rm ln\;} {|t| \over \mu ^2}
\label{diffetax}
\end{equation}
where
$\bar e ^2 = N_c (e_u ^2 + e_d ^2 + e_s ^2) = 2$ and $e_i$ are the charges of
the light quarks $u$, $d$ and $s$ respectively.

In Fig.~4a we show the differential cross-section of the process
$\gamma \gamma \rightarrow \eta_c \eta_c$.  Unlike the photoproduction of
$\eta_c$ on a nucleon target this cross-section does not vanish at $t=0$.
The result of the theoretical calculations can be conveniently represented
for $|t| < 15 \,{\rm GeV}^2$  by the exponential form:
\begin{equation}
{d\sigma \over dt} =
6.6\;{\rm fb/GeV}^2 \,{\rm exp}\, (0.167 \,{\rm GeV}^{-2}\,t).
\label{exp1}
\end{equation}
For the integrated
cross-section we get $\sigma^{tot}_{\eta_c \eta_c} = 43\,{\rm fb}$
The differential cross section for the process
$\gamma \gamma \rightarrow \eta_c + X$ is presented in Fig.~4b.
For $|t| < 15\,{\rm GeV}^2$ it can  be fitted to the exponential form:
\begin{equation}
{d\sigma \over dt} =
64\;{\rm fb/GeV}^2\,{\rm exp}\, (0.25 \,{\rm GeV}^{-2}\,t).
\label{exp2}
\end{equation}
The integrated cross-section is now
$\sigma^{tot}_{\eta_c X}(|t|>3\,{\rm GeV}^2) = 120\,{\rm fb}$.
In our calculation we set $\alpha_s (m_c^2) = 0.38$, $\tilde\alpha_s =
0.3$, $m_c = 1.4$~GeV, $m_{\eta_c}= 2.98$~GeV and
$\Gamma_{\eta_c \rightarrow \gamma\gamma}=7$~keV.

\noindent
\begin{figure}
\hbox{
\epsfxsize = 7cm
\epsfysize = 7cm 
\epsfbox[30 230 520 730]{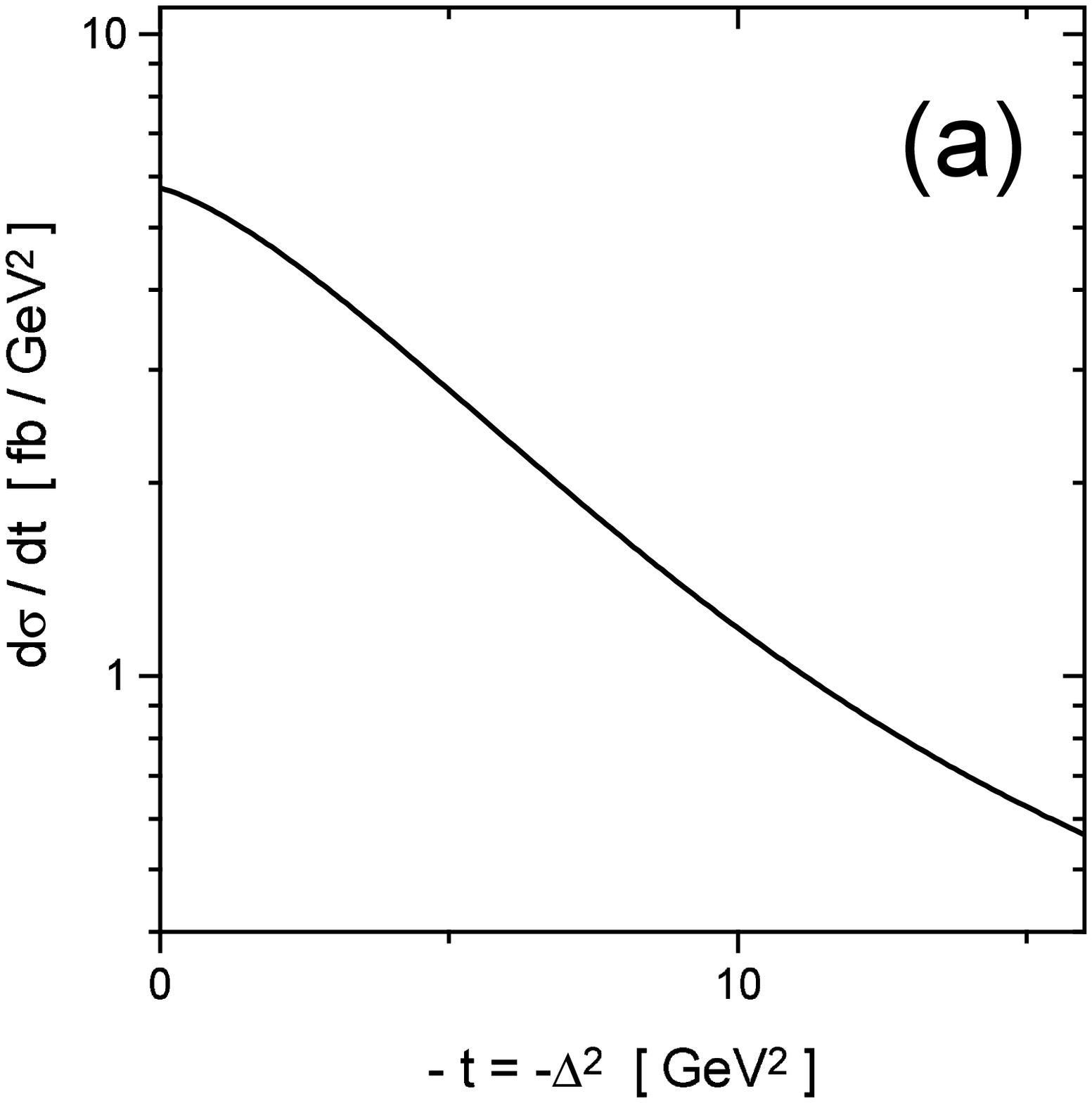}
\epsfxsize = 7cm
\epsfysize = 7cm 
\epsfbox[30 230 520 730]{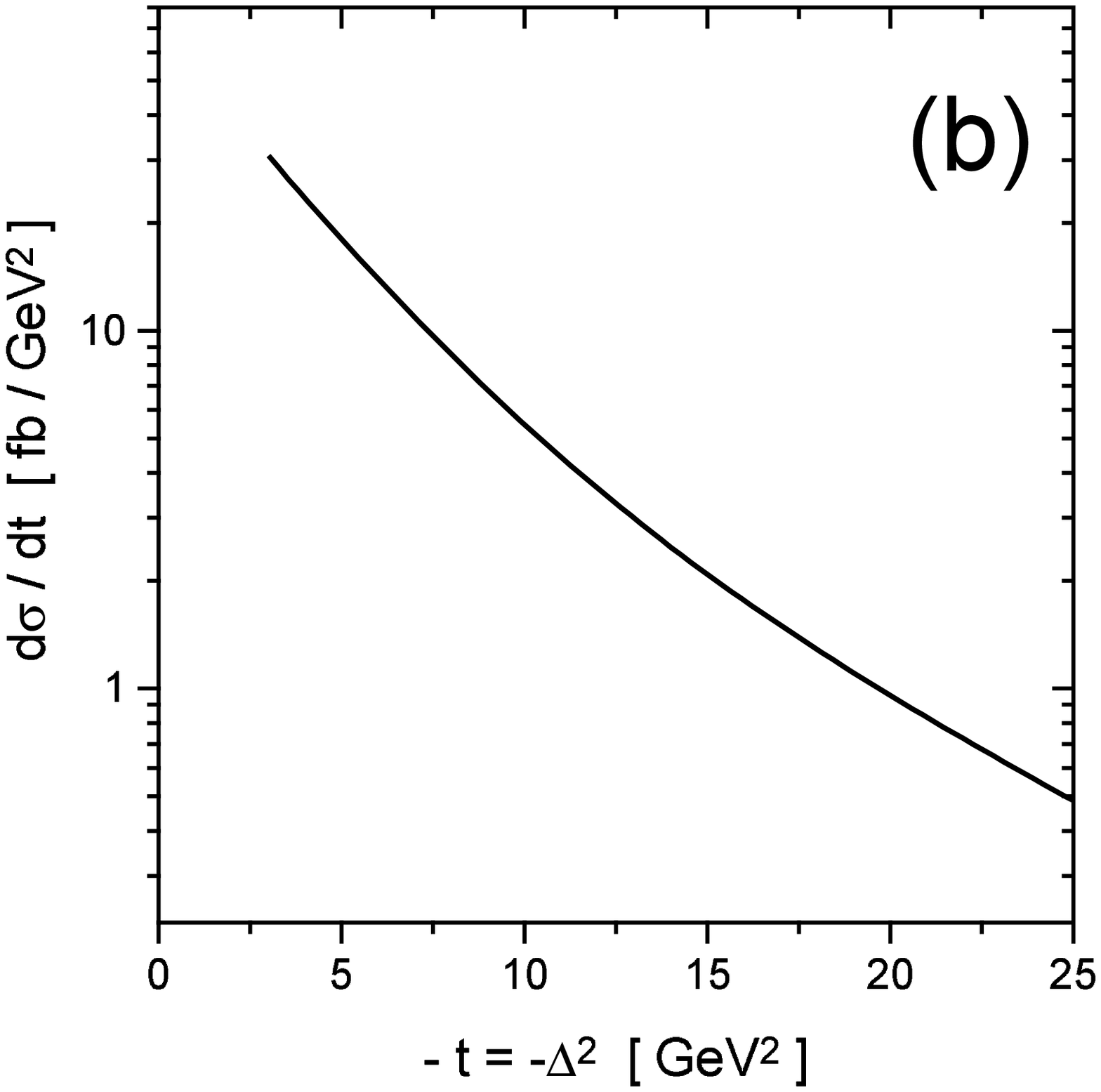}
}
\caption{
Differential cross sections of the processes 
$\gamma\gamma \rightarrow \eta_c \eta_c$ (4a) and
$\gamma\gamma \rightarrow \eta_c + X$ (4b) 
for $Q_1^2 \simeq Q_2 ^2 \simeq 0\,{\rm  GeV}^2$.}
\end{figure}

The calculated cross-sections are energy independent since they correspond
to the exchange of three elementary and non-interacting gluons.
In this approximation
the odderon singularity has its intercept $\lambda_{odd}$ equal to unity.
Interaction between gluons can boost this intercept above unity and one
can take approximately this effect into account by multiplying the
cross-sections by the enhancement factor
\begin{equation}
A_{enh}(W^2)=\bar x^{2(1-\lambda_{odd})}
\label{aenh}
\end{equation}
 where
\begin{equation}
\bar x = {M_{\eta_c}^2\over W^2}.
\label{barx}
\end{equation}
The  $\gamma \gamma$ system in $e^+e^-$ collisions has a continuous spectrum.
The C.M. energy squared $W^2$ of the $\gamma \gamma$  system is:
\begin{equation}
W^2=z_1z_2s
\label{w2}
\end{equation}
where $z_i$ are the energy fractions
of electrons (positrons) carried by the exchanged photons and $s$
denotes the C.M. energy squared of the $e^+e^-$ system.  The distribution
of those energy fractions is given by the standard flux-factors
$f_{\gamma/e}(z,Q_{min},Q_{max})$ of the virtual photons.
In the equivalent photon approximation  it is given by the following
formula \cite{YELLOW}:
\begin{equation}
f_{\gamma/e}(z,Q_{min},Q_{max})={\alpha\over 2 \pi}
\left[ {1 + (1-z)^2 \over z}\;{\rm ln}\,{Q_{max}^2\over Q_{min}^2} \right]
\label{flux}
\end{equation}
where we have neglected a small term proportional to the electron
mass squared $m_e^2$.
$Q_{min}^2$ and  $Q_{max}^2$ in eq. (\ref{flux})
denote the minimal and maximal values  of the photon virtuality.
For untagged experiments the former is given by the kinematical limit:
\begin{equation}
 Q_{min}^2 =  {m_e^2 z\over 1-z}
\label{pmin}
\end{equation}
and the latter by the antitagging condition $\theta_{e^{\pm}} < \theta_{max}$
which gives
\begin{equation}
Q_{max}^2=(1-z)E_{beam}^2 \theta_{max}^2
\label{pmax}
\end{equation}
where $\theta_{e^{\pm}}$ denotes the scattering angle of the scattered
$e^{\pm}$. Following ref.~\cite{YELLOW} we set
$\theta_{max}=30\,{\rm mrad}$.
The cross-section for the process $e^+e^- \rightarrow e^+e^- + Y$ which
for untagged $e^{\pm}$ corresponds to the production of the hadronic state
Y in the collision of almost real
(virtual) photons is given by the following
convolution integral:

$$
 \sigma_{e^+e^- \rightarrow e^+e^- + Y} =
$$
\begin{equation}
\int_0^1dz_1 \int_0^1dz_2 \Theta (W^2-W_{Y0}^2)
 \sigma_{\gamma \gamma \rightarrow Y}(W^2)
f_{\gamma/e}(z_1,Q_{min},Q_{max})f_{\gamma/e}(z_2,Q_{min},Q_{max}).
\label{conv}
\end{equation}
 In order to estimate the effective enhancement factor
due to $\lambda_{odd} > 1$ we have compared the convolution integrals:
$$
I(s,\lambda_{odd}) =
$$
\begin{equation}
 \int_0^1dz_1 \int_0^1dz_2 \Theta\left(z_{max}-
 M_{\eta_c}^2/(z_1z_2s)\right)
 f_{\gamma/e}(z_1,Q_{min},Q_{max})f_{\gamma/e}(z_2,Q_{min},Q_{max})
 \bar x^{2(1-\lambda_{odd})}
\label{is}
\end{equation}
for $\lambda_{odd}=1$ with that calculated for $\lambda_{odd}=1.07$.
We set $z_{max}=0.05$  so
that the $\gamma \gamma$ system is in the high enegy (i.e. Regge) region.
 The ratio $A=I(s,\lambda_{odd}=1.07)/I(s,\lambda_{odd}=1)$ should give
the expected enhancement factor for the given value of $s$. We get $A=1.9$
and $A=2.1$ for the LEP2 and TESLA energies respectively.  The relatively
small change in $A$ with increasing $s$ is caused by the fact that the
convolution integral is dominated by small values of $z_i$.

The magnitude of the estimated cross-sections  
$\sigma (e^+ e^- \rightarrow e^+ e^- \eta_c \eta_c) $ and  
$\sigma (e^+ e^- \rightarrow e^+ e^- \eta_c X)$ for untagged $e^+$ and 
$e^-$ in the the final state are summarized in Table~1. We give values of those
cross sections for two different C.M.~energies of incident leptons 
corresponding to LEP2 and TESLA energies and for two different values of the
odderon intercept $\lambda_{odd} = 1$ and $\lambda_{odd} = 1.07$.  
It may be seen from this table that the cross-sections are very small and so 
it may in particular be difficult to measure them with presently available 
luminosity at LEP2 \cite{YELLOW}. \\

\noindent
\begin{figure}
Table 1:The estimated cross-sections  
$\sigma (e^+ e^- \rightarrow e^+ e^- \eta_c \eta_c) $ and  
$\sigma (e^+ e^- \rightarrow e^+ e^- \eta_c X)$. \vspace{1em}

\begin{tabular}{|c|c|c|c|}
\hline
$\sqrt{s}$ [GeV] & $\lambda_{odd} - 1$ & 
$\sigma (e^+ e^- \rightarrow e^+ e^- \eta_c \eta_c)$ [fb] & 
$\sigma (e^+ e^- \rightarrow e^+ e^- \eta_c X)$ [fb] \\
\hline
180 & 0    & 1.3 & 4 \\
180 & 0.07 & 2.5 & 7 \\
500 & 0    & 3.5 & 10 \\
500 & 0.07 & 7.4 & 21 \\
\hline
\end{tabular}
\end{figure}

To sum up we have applied the formalism developed in ref.\cite{ODDRUS} to 
the quantitative analysis of the quasidiffractive processes $\gamma \gamma 
\rightarrow \eta_c \eta_c$ and $\gamma \gamma \rightarrow \eta_c X(q\bar q)$ 
within the three gluon exchange mechanism.  The main merit of those processes 
is that the corresponding cross-sections can be, in principle calculated 
within perturbative QCD.  We have estimated the corresponding 
cross-sections for the processes $e^+e^- \rightarrow e^+e^- \eta_c \eta_c$  
and $e^+e^- \rightarrow e^+e^- \eta_c X$ with untagged $e^+e^- $ which 
were found to be within the range 1---20~fb depending upon incident C.M. 
energy $\sqrt{s}$  and on the magnitude of odderon intercept.

\section*{Acknowledgments}
L.M. is grateful to J.~Czy\.zewski for inspiring discussions and to DESY Theory 
Division for hospitality.  
This research has been supported in part by the Polish Committee for Scientific
Resarch grants N0 2 P03B 89 13 and 2 P30B 044 14.

\end{document}